\documentclass[conference,twocolumn,10pt]{IEEEtran}

\pdfoutput=1
\usepackage[cmex10]{amsmath}
\usepackage{amsfonts,amssymb}
\usepackage{verbatim}

\usepackage{cite}

\ifCLASSINFOpdf
  \usepackage[pdftex]{graphicx}
  \graphicspath{{graphics/}}
  \DeclareGraphicsExtensions{.pdf,.jpeg,.png}
\else
  \usepackage[dvips]{graphicx}
  \graphicspath{{graphics/}}
  \DeclareGraphicsExtensions{.eps}
\fi

\usepackage{array}

\usepackage{mdwmath}
\usepackage{mdwtab}
\DeclareMathOperator{\erf}{erf}

\begin{document}
\bibliographystyle{IEEEtran}
%
\title{Diffusive Molecular Communication with Disruptive Flows}

\author{\IEEEauthorblockN{Adam Noel$^{\ast}$, Karen C.
Cheung$^{\ast}$, and Robert Schober$^{\ast\dagger}$}
\IEEEauthorblockA{$^{\ast}$Department of Electrical and Computer
Engineering\\
University of British Columbia, Email: \{adamn, kcheung, rschober\}@ece.ubc.ca
\\ $^{\dagger}$Institute for Digital Communications\\
Friedrich-Alexander-Universit\"{a}t Erlangen-N\"{u}rnberg (FAU), Email:
schober@LNT.de}}


\newcommand{\dbydt}[1]{\frac{d#1}{dt}}
\newcommand{\pbypx}[2]{\frac{\partial #1}{\partial #2}}
\newcommand{\psbypxs}[2]{\frac{\partial^2 #1}{\partial {#2}^2}}
\newcommand{\dbydtc}[1]{\dbydt{\conc{#1}}}
\newcommand{\thev}{\theta_v}
\newcommand{\thevi}[1]{\theta_{v#1}}
\newcommand{\theh}{\theta_h}
\newcommand{\thehi}[1]{\theta_{h#1}}
\newcommand{\x}{x}
\newcommand{\y}{y}
\newcommand{\z}{z}
\newcommand{\rad}[1]{\vec{r}_{#1}}
\newcommand{\radmag}[1]{|\rad{#1}|}


\newcommand{\kth}[1]{k_{#1}}
\newcommand{\km}{K_M}
\newcommand{\vm}{v_{max}}
\newcommand{\conc}[1]{[#1]}
\newcommand{\conco}[1]{[#1]_0}
\newcommand{\C}{C}
\newcommand{\Cx}[1]{C_{#1}}
\newcommand{\CxFun}[3]{C_{#1}(#2,#3)}
\newcommand{\Cobs}{C_{obs}}
\newcommand{\Nobs}{{\Nx{\A}}_{obs}}
\newcommand{\Nobst}[1]{\Nobs\!\left(#1\right)}
\newcommand{\Nobsn}[1]{\Nobs\left[#1\right]}
\newcommand{\Nobsavgt}{\overline{{\Nx{\A}}_{obs}}(t)}
\newcommand{\Nobsavg}[1]{\overline{{\Nx{\A}}_{obs}}\left(#1\right)}
\newcommand{\Nobsavgmax}{\overline{{\Nx{\A}}_{max}}}
\newcommand{\Nnoisetavg}[1]{\overline{{\Nx{\A}}_{n}}\left(#1\right)}
\newcommand{\Nxavg}[1]{\overline{{\Nx{\A}}_{#1}}}
\newcommand{\Nxtavg}[2]{\overline{{\Nx{\A}}_{#2}}\left(#1\right)}
\newcommand{\Nxt}[2]{{\Nx{\A}}_{#2}^\star\left(#1\right)}
\newcommand{\DMLSNntavg}[1]{\overline{{\Nx{\DMLSA}}_{n}^\star}\left(#1\right)}
\newcommand{\DMLSNnavg}{\overline{{\Nx{\DMLSA}}_{n}^\star}}
\newcommand{\DMLSNxavg}[1]{\overline{{\Nx{\DMLSA}}_{#1}^\star}}
\newcommand{\DMLSNxtavg}[2]{\overline{{\Nx{\DMLSA}}_{#2}^\star}\left(#1\right)}
\newcommand{\DMLSNxt}[2]{{\Nx{\DMLSA}}_{#2}^\star\left(#1\right)}
\newcommand{\DMLSNx}[1]{{\Nx{\DMLSA}}_{#1}^\star}
\newcommand{\Nnoiset}[1]{{\Nx{\A}}_{n}\left(#1\right)}
\newcommand{\Ntxt}[1]{{\Nx{\A}}_{TX}\left(#1\right)}
\newcommand{\Ntxtavg}[1]{\overline{{\Nx{\A}}_{TX}}\left(#1\right)}
\newcommand{\DMLSNtxtavg}[1]{\overline{{\Nx{\DMLSA}}_{tx}^\star}\left(#1\right)}
\newcommand{\DMLSNtxt}[1]{{\Nx{\DMLSA}_{tx}^\star}\left(#1\right)}
\newcommand{\DMLSNtxtavgU}[2]{\overline{{\Nx{\DMLSA}}_{tx,#2}^\star}\left(#1\right)}
\newcommand{\DMLSNtxtU}[2]{{\Nx{\DMLSA}_{tx,#2}^\star}\left(#1\right)}
\newcommand{\Cgen}{C_A(r, t)}
\newcommand{\radbind}{r_B}

\newcommand{\M}{M}
\newcommand{\smM}{m}
\newcommand{\A}{A}
\newcommand{\X}{S}
\newcommand{\vx}[1]{v_{#1}}
\newcommand{\vxvec}[1]{\vec{v}_{#1}}
\newcommand{\Pec}[1]{v^\star_{#1}}
\newcommand{\Pecper}{\Pec{\perp}}
\newcommand{\Pecpara}{\Pec{\scriptscriptstyle\parallel}}
\newcommand{\metre}{\textnormal{m}}
\newcommand{\second}{\textnormal{s}}
\newcommand{\molecule}{\textnormal{molecule}}
\newcommand{\bound}{\textnormal{bound}}
\newcommand{\argmax}{\operatornamewithlimits{argmax}}
\newcommand{\Dx}[1]{D_{#1}}
\newcommand{\Nx}[1]{N_{#1}}
\newcommand{\Nemit}{\Nx{EM}}
\newcommand{\Da}{D_\A}
\newcommand{\En}{E}
\newcommand{\en}{e}
\newcommand{\Ne}{\Nx{\En}}
\newcommand{\De}{D_\En}
\newcommand{\EA}{EA}
\newcommand{\ea}{ea}
\newcommand{\Nint}{\Nx{\EA}}
\newcommand{\Di}{D_{\EA}}
\newcommand{\Etot}{\En_{Tot}}
\newcommand{\stepl}{r_{rms}}
\newcommand{\AP}{A_P}
\newcommand{\Ri}[1]{R_{#1}}
\newcommand{\ro}{r_0}
\newcommand{\rone}{r_1}
\newcommand{\visc}{\eta}
\newcommand{\bolt}{\kth{B}}
\newcommand{\temp}{T}
\newcommand{\T}{T_{int}}
\newcommand{\Vobs}{V_{obs}}
\newcommand{\robs}{r_{obs}}
\newcommand{\Ve}{V_{enz}}
\newcommand{\tint}{\delta t}
\newcommand{\tmax}{t_{max}}
\newcommand{\Cobsfrac}{\alpha}
\newcommand{\dist}{L}
\newcommand{\DMLSA}{a}
\newcommand{\DMLSt}[1]{t_{#1}^\star}
\newcommand{\DMLSx}{x^\star}
\newcommand{\DMLSy}{y^\star}
\newcommand{\DMLSz}{z^\star}
\newcommand{\DMLSr}[1]{r_{#1}^\star}
\newcommand{\DMLSrad}[1]{\rad{#1}^\star}
\newcommand{\DMLSradmag}[1]{|\DMLSrad{#1}|}
\newcommand{\DMLSC}[1]{\Cx{#1}^\star}
\newcommand{\DMLSCxFun}[3]{{\DMLSC{#1}}(#2,#3)}
\newcommand{\DMLSc}[1]{\gamma_{#1}}
\newcommand{\DMLSV}{\Vobs^\star}
\newcommand{\DMLSNA}{\overline{{\Nx{\DMLSA}}_{obs}^\star}(\DMLSt{})}
\newcommand{\DMLSNtx}{\overline{{\Nx{\DMLSA}}_{TX}^\star}(\DMLSt{})}
\newcommand{\DMLSNAb}{\overline{{\Nx{\DMLSA}}_{obs}^\star}(\DMLSt{B})}
\newcommand{\DMLSNAmax}{{\overline{{\Nx{\DMLSA}}_{max}^\star}}}
\newcommand{\DMLStmax}[1]{{\DMLSt{#1}}_{,max}}
\newcommand{\DMLSdim}{\mathcal{D}}
\newcommand{\DMLSthreshInt}{\alpha^\star}
\newcommand{\DMLSv}[1]{v^\star_{#1}}
\newcommand{\DMLSvxvec}[1]{\vec{v}^\star_{#1}}

\newcommand{\data}[1]{W\left[#1\right]}
\newcommand{\dataSeq}{\mathbf{W}}
\newcommand{\dataSet}{\mathcal{W}}
\newcommand{\dataObs}[1]{\hat{W}\left[#1\right]}
\newcommand{\numX}[2]{n_{#1}\left(#2\right)}
\newcommand{\thresh}{\xi}
\newcommand{\poissBar}{\Big|_\textnormal{Poiss}}
\newcommand{\gaussBar}{\Big|_\textnormal{Gauss}}
\newcommand{\eqBar}[2]{\Big|_{#1 = #2}}
\newcommand{\Pobs}{P_{obs}}
\newcommand{\Pobsx}[1]{P_{obs}\left(#1\right)}
\newcommand{\Pone}{P_1}
\newcommand{\Pzero}{P_0}
\newcommand{\Pe}[1]{P_{err}\left[#1\right]}
\newcommand{\Peavg}[1]{\overline{P_{err}}\left[#1\right]}
\newcommand{\threshInterval}{\alpha}
\newcommand{\pstay}[1]{P_{stay}\left(#1\right)}
\newcommand{\pleave}[1]{P_{leave}\left(#1\right)}
\newcommand{\parrive}[1]{P_{arr}\left(#1\right)}

\newcommand{\VAmem}{F}
\newcommand{\VAstate}{f}
\newcommand{\VAdataObs}[2]{\hat{W}_{\VAstate_{#2}}\left[#1\right]}
\newcommand{\VAcurLL}[2]{\Phi_{\VAstate_{#2}}\left[#1\right]}
\newcommand{\VAcumLL}[2]{L_{\VAstate_{#2}}\left[#1\right]}

\newcommand{\weight}[1]{w_{#1}}

\newcommand{\fof}[1]{f\left(#1\right)}
\newcommand{\floor}[1]{\lfloor#1\rfloor}
\newcommand{\lam}[1]{W\left(#1\right)}
\newcommand{\EXP}[1]{\exp\left(#1\right)}
\newcommand{\ERF}[1]{\erf\left(#1\right)}
\newcommand{\SIN}[1]{\sin\left(#1\right)}
\newcommand{\SINH}[1]{\sinh\left(#1\right)}
\newcommand{\COS}[1]{\cos\left(#1\right)}
\newcommand{\COSH}[1]{\cosh\left(#1\right)}
\newcommand{\Ix}[2]{I_{#1}\!\left(#2\right)}
\newcommand{\Jx}[2]{J_{#1}\!\left(#2\right)}
\newcommand{\E}[1]{E\left[#1\right]}
\newcommand{\GamFcn}[1]{\Gamma\!\left(#1\right)}
\newcommand{\mean}[1]{\mu_{#1}}
\newcommand{\var}[1]{\sigma_{#1}^2}

\newcommand{\B}[1]{B_{#1}}
\newcommand{\w}{w}
\newcommand{\n}{n}
\newcommand{\gx}[1]{g\left(#1\right)}
\newcommand{\hx}[1]{h\left(#1\right)}
\newcommand{\tx}[1]{t\left(#1\right)}
\newcommand{\ux}[1]{u\left(#1\right)}
\newcommand{\deltObs}{t_{o}}
\newcommand{\sx}[1]{s_{#1}}

\newcommand{\new}[1]{\textbf{#1}}
\newcommand{\ISI}{ISI}
\newcommand{\DDFSE}{DDFSE}
\newcommand{\PDF}{PDF}
\newcommand{\CDF}{CDF}
\newcommand{\AWGN}{AWGN}

\newtheorem{theorem}{Theorem}

\maketitle

\begin{abstract}
In this paper, we study the performance of detectors
in a diffusive molecular communication environment
where steady uniform flow is present. We derive the
expected number of information molecules to be
observed in a passive spherical receiver, and determine
the impact of flow on the assumption that the concentration
of molecules throughout the receiver is uniform.
Simulation results show the impact of advection on detector
performance as a function
of the flow's magnitude and direction.
We highlight that there are disruptive flows, i.e., flows
that are not in the direction of information transmission,
that lead to an improvement in detector performance as long as
the disruptive flow does not dominate diffusion and sufficient
samples are taken.
\end{abstract}

\section{Introduction}

Future applications of nanotechnology will be
enabled by the ability of individual devices
with nanoscale components to communicate amongst
themselves and share information. Fields that
require diagnostics or actions on a small
scale, such as healthcare and manufacturing,
are envisioned to benefit from the deployment
of these connected devices, collectively known as nanonetworks;
see \cite{RefWorks:540, RefWorks:608}.
Molecular communication is a physical layer design strategy
for nanonetworks where transmitters release molecules
that freely travel or are carried to their intended destinations.
This strategy is bio-inspired; cellular systems
use molecules to communicate via integration with
biochemical mechanisms; see \cite{RefWorks:750}.

The simplest molecular transport mechanism is free diffusion, where
molecules move randomly due to collisions with other molecules
in the environment. No external energy is required, unlike in
active propagation methods where energy is consumed to direct
molecules towards the receiver. Furthermore, networks of
devices communicating via diffusion can form spontaneously
because there are no fixed connections between devices.
However, communication via diffusion alone is limited by
both propagation time and reliability as the distance between
devices increases. In a purely diffusive environment,
intersymbol interference (\ISI) is a major problem
because the receiver cannot differentiate between the arrivals
of the same type of molecule emitted at different times.

Flows play an important role in many diffusive environments
where molecular communication networks may be deployed,
especially when the distance that
molecules must travel is greater than what can
be practically achieved via diffusion alone.
For example, the advection of blood in the body enables
the transport of oxygen from the lungs to tissues
and also facilitates the removal of waste and toxic
molecules via the liver and kidneys; see \cite{RefWorks:750}.
Flows can also help mitigate \ISI~in a communications context
by carrying ``old''
molecules away from the receiver.

Analytically, the simplest flow is both steady and uniform.
In a steady flow, the pressure, density, and velocity components at
each point in the stream do not change with time; see \cite{RefWorks:751}.
If uniform, then these components are identical throughout the
environment of interest. Existing literature in molecular
communication has generally assumed that flow is not only both steady
and uniform (which are not necessarily realistic assumptions),
but also only in the direction
of transmission, i.e., in the same direction as a line pointing
from the transmitter to its intended receiver, cf. e.g. \cite{RefWorks:513,
RefWorks:607,RefWorks:615}.
We define \emph{disruptive} flows as any flow component
that is not in the direction of transmission. These
flows are literally destructive in that they reduce the
peak number of molecules expected to be observed at the receiver.

In \cite{RefWorks:747}, we studied receiver detection schemes for
a 3-dimensional stochastic diffusion environment between a single
transmitter and receiver. Our physical model included steady
uniform flow in any direction (a notion that was also recently
considered in a 2-dimensional environment in \cite{RefWorks:734}).
Simulation results showed the
average detector bit error probabilities
under a few sample cases of flow. We observed
that there were disruptive flows under which both optimal and suboptimal
detectors could \emph{improve} over the no-flow case, because
the advective removal of unintended (\ISI) molecules mitigated the
removal of intended molecules. These results
suggested that bi-directional transmission in flowing environments
is not only possible but could actually be better than in an environment
without flow.
Thus, it is of interest to investigate precisely under what conditions
advection is beneficial to diffusive communication, i.e., what magnitudes
and directions of flow decrease the probability of bit errors
for which detectors, and when advection
degrades communication.

Furthermore, we assumed for tractability of analysis
in \cite{RefWorks:747} that the expected concentration
of information molecules throughout the receiver was \emph{equal} to
that expected at the center of the receiver.
We hereafter refer to this assumption as the \emph{uniform concentration
assumption}. We assessed this assumption's accuracy in environments without flow
in \cite{RefWorks:706}, but evaluating its accuracy in
flowing environments has been an open problem.

This paper extends the consideration of steady uniform flow in
\cite{RefWorks:747} to make the following contributions:
\begin{enumerate}
    \item We derive the expected number of information molecules
    at an ideal spherical receiver due to an emission by a point source
    into an unbounded diffusive environment with a steady uniform flow.
    The derivation is made \emph{without} the uniform concentration assumption
    and a closed-form solution is possible only when the
    flow is parallel to a line joining the source and
    receiver.
    \item We assess the uniform concentration assumption by measuring
    the relative deviation in the expected concentration of information
    molecules when the assumption is applied.
    This comparison is made
    using dimensionless values, as we used in \cite{RefWorks:706},
    so that the results scale to any reference dimension.
    \item We compare the performance of the optimal and weighted
    sum detectors described in \cite{RefWorks:747} under a wide
    range of steady uniform flows.
    We discuss the types of flow
    for which detector performance is improved over the no-flow case,
    highlighting disruptive flows that enable a decrease in the
    probability of error. 
\end{enumerate}

We note that, in \cite{RefWorks:747}, we also considered the presence
of enzyme molecules in the propagation environment that are
capable of degrading the information molecules as a strategy
to mitigate \ISI. For clarity of exposition, and
because advection can also mitigate \ISI, we do not
include the effect of enzymes in the model presented in this paper.
However, all results presented here can be easily extended to
the case where enzymes are in the propagation environment.


The rest of this paper is organized as follows.
The physical environment is described in dimensional and dimensionless
forms and the detectors are summarized in Section~\ref{sec_model}.
In Section~\ref{sec_obs}, we derive the expected number of molecules
observed at the receiver, both with and without the uniform concentration
assumption.
In Section~\ref{sec_det}, we summarize the optimal and weighted
sum detectors that we presented in \cite{RefWorks:747}.
Numerical and simulation results
are presented in Section~\ref{sec_results}. We draw our
conclusions in Section~\ref{sec_concl}.

\section{System Model}
\label{sec_model}

In this section, we describe the system model that we considered in
\cite{RefWorks:747} and then, unlike in \cite{RefWorks:747},
translate it into dimensionless form.

\subsection{Dimensional Form}

The receiver is a sphere with radius $\robs$ and
volume $\Vobs$ that is fixed and centered at the origin of an infinite,
3-dimensional aqueous environment of constant uniform temperature
and viscosity. The transmitter is fixed at location $\{-\x_0,0,0\}$
(without loss of generality). By symmetry, the concentrations observed in
this environment are equivalent (by a factor of 2)
to those in the semi-infinite
case where $\z \ge 0$ is the aqueous environment, the $\x\y$-plane is an elastic
boundary, and the receiver is a hemisphere whose circle face lies on the boundary;
see \cite[Eq. 2.7]{RefWorks:586}.
This equivalent environment could describe, for example, a small transmitter
and receiver mounted along the wall of a large blood vessel or artery.
The receiver is a (virtual) passive observer
that does not impede diffusion or initiate chemical reactions (this
assumption focuses our analysis on the impact of the propagation
environment and enables tractability).
A steady uniform flow (or drift) exists and is
defined by its velocity component along each dimension, i.e.,
$\vxvec{} = \{\vx{\x},\vx{\y},\vx{\z}\}$. The placement of the transmitter
is so that a positive $\vx{\x}$ is in the direction
of the receiver from the transmitter. Thus, a negative $\vx{\x}$,
a nonzero $\vx{\y}$, or a nonzero $\vx{\z}$ represent disruptive flows.

The transmitter has a binary sequence of length $\B{}$,
$\dataSeq = \{\data{1},\data{2},\dots,\data{\B{}}\}$, to
send to the receiver, where $\data{j}$ is the $j$th information bit and
$\Pr(\data{j} = 1) = \Pone$. The transmitter
uses binary modulation and transmission intervals of duration $\T$ seconds. To send
a binary $1$, $\Nemit$ $\A$ molecules are released at the start of the bit interval.
To send a binary $0$, no molecules are released.

We assume that the environment has additional sources
of $\A$ molecules, for example via interference from other communication links
or via some other chemical process that generates $\A$ molecules. These external
noise sources are distinct from the diffusion of $\A$ molecules emitted by
the transmitter. As in \cite{RefWorks:747}, we only assume that
we know the cumulative impact of all noise sources on the received signal, and
that we can characterize this impact as a Poisson random variable with time-varying
mean $\Nnoisetavg{t}$.
This noise model is sufficient to model either of the example cases above.
However, we omit the impact of steady uniform flow on the noise sources themselves
(since we do not specify the origin of the noise), and
we will only consider additive noise sources with constant mean in our
simulations. We leave a more precise characterization of noise and
interference for future work.

The concentration of
$\A$ molecules at the point defined by vector $\rad{}$ and at time $t$ in
$\molecule\cdot\metre^{-3}$ is $\CxFun{\A}{\rad{}}{t}$ (or written as
$\Cx{\A}$ for compactness). We assume that these molecules diffuse
independently once they are released by the
transmitter or noise sources. In addition, due to the constant uniform
temperature and viscosity of the environment,
the $\A$ molecules diffuse with constant diffusion coefficient $\Dx{\A}$.
The differential equation describing the motion of $\A$ molecules
due to both diffusion and advection is
a modified version of Fick's second law, written as
\cite[Ch. 4]{RefWorks:630}
\begin{equation}
\label{JUN12_33}
\pbypx{\Cx{\A}}{t} = \Dx{\A}\nabla^2\Cx{\A} - \vx{\x}\pbypx{\Cx{\A}}{\x}
- \vx{\y}\pbypx{\Cx{\A}}{\y}
- \vx{\z}\pbypx{\Cx{\A}}{\z},
\end{equation}
which can be easily solved to find the \emph{expected} point concentration due
to an emission of $\Nemit$ molecules by the transmitter at time
$t = 0$. It is easy
to show with a moving reference frame that this expected concentration
at point $\{\x,\y,\z\}$ is
\begin{equation}
\label{JUN12_47}
\Cx{\A} = \frac{\Nemit}{(4\pi \Da
t)^{3/2}}\EXP{- \frac{\radmag{}^2}{4\Da t}},
\end{equation}
where
$\radmag{}^2 = (\x + \x_0 - \vx{\x}t)^2 + (\y - \vx{\y}t)^2 + (\z - \vx{\z}t)^2$
is the square of the \emph{effective} distance from
the transmitter at $\{-\x_0,0,0\}$
to $\{\x,\y,\z\}$.

\subsection{Dimensionless Form}

For dimensional analysis, reference variables are used to scale all
parameters into dimensionless form; please refer to
\cite{RefWorks:633} for more on dimensional analysis.
As in \cite{RefWorks:706}, we define
reference distance $\dist$ in $\metre$ and reference number
of molecules $\Nemit$ (i.e., the transmitter releases one
dimenionless molecule to send a binary $1$). We also define
reference concentration $\Cx{0} = \Nemit/\dist^3$ in
$\molecule\cdot\metre^{-3}$.
We then define the dimensionless concentration of $\A$ molecules
as $\DMLSC{\DMLSA} = \Cx{\A}/\Cx{0}$, dimensionless time
as $\DMLSt{} = \Dx{\A}t/\dist^2$,
and dimensionless coordinates along the three axes as
\begin{equation}
\label{AUG12_43_coor}
\DMLSx = \frac{\x}{\dist}, \quad
\DMLSy = \frac{\y}{\dist}, \quad
\DMLSz = \frac{\z}{\dist},
\end{equation}
such that variables with
a ``$\star$'' superscript
are equal to the corresponding dimensional variables scaled by the
appropriate reference variables. Advection is represented
dimensionlessly with the Peclet number,
$\Pec{}$, written as \cite[Eq. 1.3.1]{RefWorks:750}
\begin{equation}
\label{EQ13_07_30}
\Pec{} = \frac{\vx{}\dist}{\Dx{\A}},
\end{equation}
where $\vx{} = |\vxvec{}|$ is the speed of the fluid. $\Pec{}$
measures the relative impact of advection versus diffusion
on molecular transport.
If $\Pec{} \ll 1$, i.e., if $\vx{} \to 0$,
then diffusion dominates particle motion and advection
can be ignored. If
$\Pec{} \gg 1$, then advection dominates particle motion and
diffusion can be ignored. Under advection-dominant motion,
disruptive flows should prevent successful communication (no matter what detector
is used) and non-disruptive flows should aid communication (as long as the
receiver takes a sample while molecules are flowing
through it). Thus, the impact of
steady uniform flow on communication depends on the
\emph{direction} of flow, and it is important to consider $\Pec{}$-values
much less and much greater than $1$. We define $\Pec{}$ along each
dimension as
\begin{equation}
\label{EQ13_07_29}
\Pecpara = \frac{\vx{\x}\dist}{\Dx{\A}}, \quad
\Pec{\perp,1} = \frac{\vx{\y}\dist}{\Dx{\A}}, \quad
\Pec{\perp,2} = \frac{\vx{\z}\dist}{\Dx{\A}},
\end{equation}
but we note that, without loss of generality (because the
transmitter is on the $\x$-axis), we can set $\Pec{\perp,2} = 0$
and write $\Pec{\perp,1} = \Pecper$.
Given the dimensionless model, (\ref{JUN12_33}) becomes
\begin{equation}
\label{JUN12_33_DMLS}
\pbypx{\DMLSC{\DMLSA}}{\DMLSt{}} = \nabla^2\DMLSC{\DMLSA}
- \Pecpara\pbypx{\DMLSC{\DMLSA}}{\DMLSx}
- \Pecper\pbypx{\DMLSC{\DMLSA}}{\DMLSy},
\end{equation}
where
\begin{align}
\pbypx{\DMLSC{\DMLSA}}{\DMLSt{}} =
&\;\pbypx{\Cx{\A}}{t}\frac{\dist^2}{\Dx{\A}\Cx{0}}, \quad
\label{AUG12_45}
\nabla^2\DMLSC{\DMLSA} = \frac{\dist^2}{\Cx{0}}\nabla^2\Cx{\A},
\\
\label{AUG12_45_v}
\pbypx{\DMLSC{\DMLSA}}{\DMLSx} = &\;
\pbypx{\Cx{\A}}{\x}\frac{\dist}{\Cx{0}}, \quad
\pbypx{\DMLSC{\DMLSA}}{\DMLSy} = \pbypx{\Cx{\A}}{\y}\frac{\dist}{\Cx{0}},
\end{align}
and (\ref{JUN12_47}) becomes
\begin{equation}
\label{APR12_22}
\DMLSC{\DMLSA} = \frac{1}{(4\pi
\DMLSt{})^{3/2}}\EXP{\frac{-\DMLSradmag{}^2}{4 \DMLSt{}}},
\end{equation}
where
$\DMLSradmag{}^2 = (\DMLSx + \DMLSx_0 - \Pecpara\DMLSt{})^2 +
(\DMLSy - \Pecper\DMLSt{})^2 + {\DMLSz}^2$
is the square of the effective distance from the transmitter at
$\{-\DMLSx_0,0,0\}$ to $\{\DMLSx,\DMLSy,\DMLSz\}$.

\section{Receiver Signal}
\label{sec_obs}

In this section, we derive the expected dimensionless number
of molecules observed at the receiver that were emitted by
the transmitter at $\DMLSt{} = 0$, $\DMLSNxtavg{\DMLSt{}}{0}$.
This derivation enables
comparison with the uniform concentration assumption, i.e., the
assumption that the expected concentration
of information molecules throughout the receiver is equal to
that expected at the center of the receiver.
Then, we write the general time-varying receiver signal
$\Nobst{t}$ as a function of the transmitter's emissions and
the external noise sources.

The receiver is a passive observer, so
the expected dimensionless number of $\A$ molecules within the
dimensionless receiver volume is found by integrating (\ref{APR12_22})
over $\DMLSV$, i.e.,
\begin{equation}
\label{APR12_42_DMLS}
\DMLSNxtavg{\DMLSt{}}{0} = \int\limits_0^{\DMLSr{obs}}
\int\limits_{0}^{2\pi}
\int\limits_{0}^{\pi}
\DMLSC{\DMLSA}{\DMLSr{i}}^2\sin\theta
d\theta d\phi d\DMLSr{i},
\end{equation}
where $\DMLSr{i}$ is the magnitude of the distance from
the origin (i.e., the center of the receiver) to the arbitrary point
$\{\DMLSx,\DMLSy,\DMLSz\}$ within $\DMLSV$.
The first step in solving (\ref{APR12_42_DMLS}) is to convert
$\DMLSradmag{}^2$ in (\ref{APR12_22}) from Cartesian
to spherical coordinates.
It is straightforward to show that
\begin{align}
\DMLSradmag{}^2 = &\;
{\DMLSr{i}}^2 + {\DMLSx_0}^2 - 2\DMLSt{}\DMLSx_0\Pecpara
- 2\DMLSx_0\DMLSr{i}\cos\phi\sin\theta
\nonumber \\
& -2\DMLSt{}\DMLSr{i}
\left(\Pecpara\cos\phi\sin\theta + \Pecper\sin\phi\sin\theta\right)
\nonumber \\
&  + {\DMLSt{}}^2\left({\Pecpara}^2
+ {\Pecper}^2\right),
\label{EQ13_07_25}
\end{align}
where $\phi = \tan^{-1}\left(\DMLSy/\DMLSx\right)$ and $\theta =
\cos^{-1}\left(\DMLSz/\DMLSr{i}\right)$.
Generally, (\ref{APR12_42_DMLS}) does not have a known closed-form
solution, due to the sum of trigonometric terms in the exponential
in (\ref{APR12_22}).
We can integrate (\ref{APR12_42_DMLS}) over $\DMLSr{i}$ using
substitution, integration by parts, the definition of the error
function, i.e., \cite[Eq. 3.1.1]{RefWorks:700}
\begin{equation}
\label{APR12_32}
\ERF{a} = \frac{2}{\pi^\frac{1}{2}}\int_0^a \EXP{-b^2} db,
\end{equation}
and the integral \cite[Eq. 4.1.4]{RefWorks:700}
\begin{equation}
\int a\ERF{a}da = \frac{1}{2}\ERF{a}\left(a^2-\frac{1}{2}\right)+
\frac{a}{2\pi^\frac{1}{2}}\EXP{-a^2}.
\end{equation}
It can then be shown that
$\DMLSNxtavg{\DMLSt{}}{0}$ becomes the integration of
\begin{align}
& (2\pi^\frac{3}{2})^{-1}
\EXP{\beta_1^2\DMLSt{} + \beta_2}\sin\theta
\Big[\beta_1{\DMLSt{}}^\frac{1}{2}
\EXP{-\beta_1^2{\DMLSt{}}} + \pi^\frac{1}{2}/2
\nonumber \\
& \times \left(1+2\beta_1^2{\DMLSt{}}\right)\!\!
\left(\ERF{\DMLSr{obs}{\DMLSt{}}^{-\frac{1}{2}}/2-
\beta_1{\DMLSt{}}^\frac{1}{2}} \!-
\ERF{-\beta_1{\DMLSt{}}^\frac{1}{2}}\right) \nonumber \\
& -\left(\beta_1{\DMLSt{}}^\frac{1}{2} + \DMLSr{obs}{\DMLSt{}}^{-\frac{1}{2}}/2\right)
\EXP{-(\DMLSr{obs}{\DMLSt{}}^{-\frac{1}{2}}/2 - \beta_1{\DMLSt{}}^\frac{1}{2})^2}
\Big],
\label{APR12_42_DMLS_r}
\end{align}
over $\theta \in [0,\pi]$ and $\phi \in [0,2\pi]$, where
\begin{align}
\beta_1 = &\; \frac{\sin\theta}{2}\left(\Pecpara\cos\phi
+\Pecper\sin\phi + \frac{\DMLSx_0}{\DMLSt{}}\cos\phi\right),
\\
\beta_2 = &\; \frac{\DMLSx_0\Pecpara}{2} -
\frac{\DMLSt{}}{4}\left({\Pecpara}^2 + {\Pecper}^2\right) -
\frac{{\DMLSx_0}^2}{4\DMLSt{}}.
\end{align}

For general steady uniform flow $\Pec{} \neq 0$,
(\ref{APR12_42_DMLS}) can be solved by numerically
integrating (\ref{APR12_42_DMLS_r}) over $\theta$ and $\phi$.
However, for the special
case $\Pecper = 0$, such that the direction of flow
is parallel to a line between the source and receiver,
we can apply Theorem 2 in \cite{RefWorks:706} using
a change of variables and solve (\ref{APR12_42_DMLS}) as
\begin{align}
\DMLSNxtavg{\DMLSt{}}{0} = &\; \frac{1}{2}\left[\ERF{\frac{\DMLSr{obs}\!-
\DMLSrad{eff}}{2{\DMLSt{}}^\frac{1}{2}}} +
\ERF{\frac{\DMLSr{obs}\!+\DMLSrad{eff}}{2{\DMLSt{}}^\frac{1}{2}}}\right] \nonumber
\\
& +
\frac{1}{\DMLSrad{eff}}\sqrt{\frac{\DMLSt{}}{\pi}}
\Bigg[\EXP{-\frac{(\DMLSrad{eff}+\DMLSr{obs})^2}{4\DMLSt{}}} \nonumber \\
& - \EXP{-\frac{(\DMLSrad{eff}-\DMLSr{obs})^2}{4\DMLSt{}}}\Bigg],
\label{EQ13_07_33}
\end{align}
where $\DMLSrad{eff} = -\left(\DMLSx_0 - \Pecpara\DMLSt{}\right)$ is
the effective distance along the $\DMLSx$-axis
from the transmitter to the center of the receiver.

If we apply the uniform concentration assumption, then the evaluation
of (\ref{APR12_42_DMLS}) is simply the product of $\DMLSV$ and
$\DMLSC{\DMLSA}$ at the center of the receiver, i.e., at the origin.
Thus, we have
\begin{equation}
\label{EQ13_07_26_DMLS}
\DMLSNxtavg{\DMLSt{}}{0} = \frac{\DMLSV}{(4\pi
\DMLSt{})^{3/2}}\EXP{-\frac{{\DMLSrad{eff}}{}^2}{4 \DMLSt{}}
- \frac{{\DMLSt{}}{\Pecper}^2}{4}},
\end{equation}
and in Section~\ref{sec_results} we will measure the relative
deviation of (\ref{EQ13_07_26_DMLS}) from (\ref{APR12_42_DMLS}),
where (\ref{APR12_42_DMLS}) is solved using (\ref{EQ13_07_33})
if $\Pecper = 0$ and by numerically integrating (\ref{APR12_42_DMLS_r})
otherwise.

It is straightforward to derive the statistics of the
general receiver signal $\Nobst{t}$ based on $\DMLSNxtavg{\DMLSt{}}{0}$
and the transmitter sequence $\dataSeq$. Assuming constant ideal diffusion, then
the behavior of individual $\A$ molecules is independent,
and $\Nobst{t}$ is a sum of time-varying Poisson random
variables (as described in \cite{RefWorks:747}),
with time-varying mean
\begin{equation}
\Nobsavgt = \Ntxtavg{t} + \Nnoisetavg{t},
\label{EQ13_05_28}
\end{equation}
where $\Nnoisetavg{t}$ is the mean number of molecules from the noise
sources, $\Ntxtavg{t}$ is the mean number of molecules from
emissions by the transmitter, i.e.,
\begin{equation}
\label{EQ13_04_02}
\Ntxtavg{t} =
\sum_{j=1}^{\floor{\frac{t}{\T}+1}}\data{j}\Nxtavg{t-(j-1)\T}{0},
\end{equation}
and $\Nxtavg{t}{0}$ is $\DMLSNxtavg{\DMLSt{}}{0}$ in dimensional form. 

\section{Detector Summary}
\label{sec_det}

To study the impact of advection on detector performance, we
implement the detectors that we proposed in \cite{RefWorks:747}.
In this section, we summarize these detectors.
The detectors rely on a common sampling scheme, where the receiver
makes $\M$ observations in every bit interval, and we assume in this paper
that the observations are independent. The value of the $\smM$th
observation in the $j$th interval is labeled $\sx{j,\smM}$.
We define the sampling times within a single interval as the
function $\gx{\smM}$, and the global time sampling function
$\tx{j,\smM} = j\T + \gx{\smM}$, where
$j = \{1,2,\ldots,\B{}\}, \smM = \{1,2,\ldots,\M\}$.
We assume that the transmitter and receiver are perfectly synchronized.
Synchronization amongst devices in a non-advective environment
was achieved in \cite{RefWorks:773}
via the diffusion of inhibitory molecules. However, synchronization in flowing
environments remains an open problem.

We use the maximum likelihood optimal sequence detector to
give a lower bound on the bit error probability.
The optimal receiver decision rule, in a maximum likelihood sense,
is to select the most likely sequence $\dataObs{j}$
given the joint likelihood of all received samples, i.e.,
\begin{equation}
\label{FEB13_03}
\dataObs{j}\eqBar{j}{\{1,2,\ldots,\B{}\}} =
\argmax_{\data{j}, j = \{1,2,\ldots,\B{}\}} \Pr\left(\Nobs\right)
\end{equation}
where, assuming independent samples,
\begin{equation}
\label{FEB13_21}
\Pr(\Nobs) = \prod_{j=1}^{\B{}}\prod_{\smM=1}^\M
\Pr\big(\Nobst{\tx{j,\smM}} = \sx{j,\smM} \mid \dataSeq\big),
\end{equation}
and the individual likelihoods can easily be found by recognizing
that $\Nobst{t}$ is a Poisson random variable, as discussed in
Section~\ref{sec_obs}.
The complexity can be reduced by applying
methods such as the modified Viterbi algorithm that we proposed
in \cite{RefWorks:747}, where we limit the explicit channel
memory to $\VAmem$ prior bit intervals.
In the simulations presented in
Section~\ref{sec_results}, we choose $\VAmem = 2$ as a compromise between
computational complexity and observable error probability
(theoretically, the channel memory of a diffusive environment is infinite).

Weighted sum detectors have considerably less complexity than the optimal
sequence detector, enable a tractable derivation of the bit error
probability, and are more suitable for practical use; neurons combine inputs
from synapses using a weighted sum detector (see \cite[Ch. 12]{RefWorks:587}).
The decision rule of the weighted sum detector in the $j$th bit interval is
\begin{equation}
\dataObs{j} = \left\{
 \begin{array}{rl}
  1 & \text{if} \quad
  \sum_{\smM = 1}^{\M}\weight{\smM}\Nobst{\tx{j,\smM}} \ge \thresh,\\
  0 & \text{otherwise},
 \end{array} \right.
\label{FEB13_33}
\end{equation}
where $\weight{\smM}$ is the weight of the $\smM$th observation and
$\thresh$ is the binary decision threshold. The selection of $\thresh$
is outside the scope of this work and we assume that the optimal
$\thresh$ for the given environment is found via numerical search.

We consider two special cases of weighted sum detectors.
First, the equal weight detector, where each weight is equal
to $1$, is the simplest weighted sum detector. Second, the
matched filter detector, which we showed via simulation in
\cite{RefWorks:747} to achieve performance equal to that of the optimal
sequence detector in the absence of \ISI~(even though the noise that we
consider is Poisson-distributed and not Gaussian), has weights based on
the number of molecules expected due to an emission by the transmitter in the
current bit interval. Further details on
the evaluation of the expected bit error probabilities of these
weighted sum detectors can be found in \cite{RefWorks:747}.

\section{Numerical Results}
\label{sec_results}

In this section, we measure the deviation in the uniform concentration
assumption for $\DMLSNxtavg{\DMLSt{}}{0}$ over time as a
function of the receiver's distance
from the transmitter and of the steady uniform flow present in the
environment. Then, we assess detector performance of a sequence
of bits under a range of steady uniform flows.

\subsection{Uniform Concentration Assumption}

We set the reference distance $\dist = \x_0$. A reference
$\Nemit$ is not needed because we measure relative
deviation in concentration.
Based on our observations in \cite{RefWorks:706}, we set
$\DMLSr{obs} = 0.1$ so that the deviation of the uniform concentration
assumption in the no-flow case is less than
$1\,\%$ for all $\DMLSt{} > 0.1$.
The maximum number of molecules in the no-flow case
is expected at $\DMLSt{} = \frac{1}{6}$.
We separately vary $\Pecpara$ and
$\Pecper$ because any flow is equivalent
to a combination of $\Pecpara$ and $\Pecper$.

\begin{figure}[!tb]
\centering
\includegraphics[width=\linewidth]{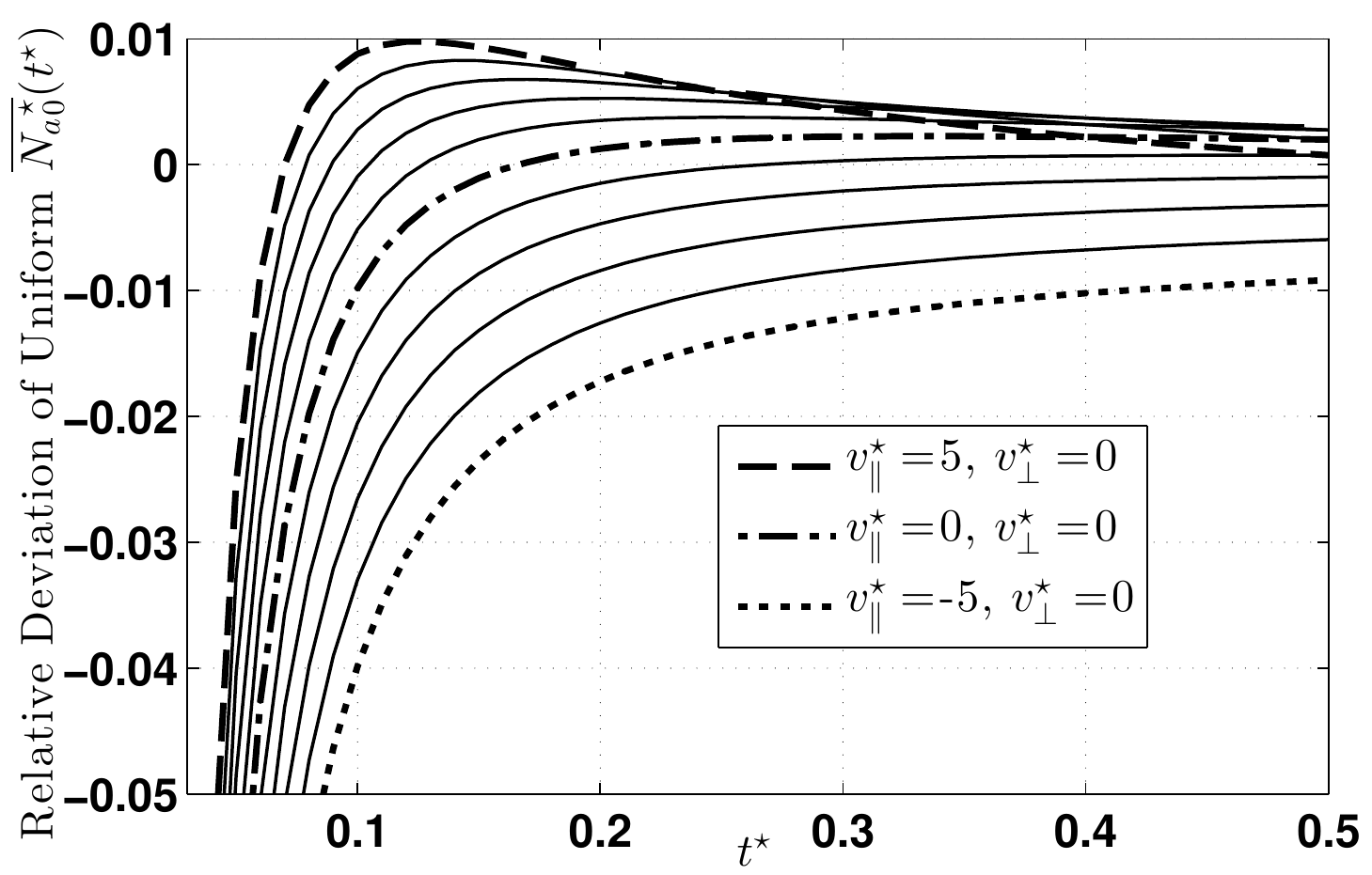}
\caption{The relative deviation in $\DMLSNxtavg{\DMLSt{}}{0}$ from the true value
(\ref{APR12_42_DMLS}) at the receiver when the uniform concentration
assumption (\ref{EQ13_07_26_DMLS}) is applied.
The flow $\Pecpara$ is varied from $-5$
to $5$ in increments of $1$.}
\label{uniform_vx}
\end{figure}

In Fig.~\ref{uniform_vx}, we assess the uniform concentration assumption
over time while varying $\Pecpara$ from $-5$ to $5$ in increments
of $1$. All flows severely
underestimate $\DMLSNxtavg{\DMLSt{}}{0}$ (i.e., deviation is much less than 0)
for $\DMLSt{} < 0.05$. This underestimation is because molecules
are expected to reach the edge of $\DMLSV$ (and
thus be observed) before they are expected at the center.
When $\Pecpara$ is positive, $\DMLSNxtavg{\DMLSt{}}{0}$
is overestimated earlier than in the no-flow case because the
peak number of molecules is observed sooner and the center of
$\DMLSV$ is closer to the transmitter than most of $\DMLSV$.
When $\Pecpara$ is negative, $\DMLSNxtavg{\DMLSt{}}{0}$ is
underestimated longer than in the no-flow case.
Importantly, applying the uniform concentration assumption to
all degrees of flow within the range
$-2 \le \Pecpara \le 4$ (for which advection is not dominant, as
we will see in the next subsection) 
introduces a deviation of less than $2\,\%$ for all $\DMLSt{} > 0.1$.

\begin{figure}[!tb]
\centering
\includegraphics[width=\linewidth]{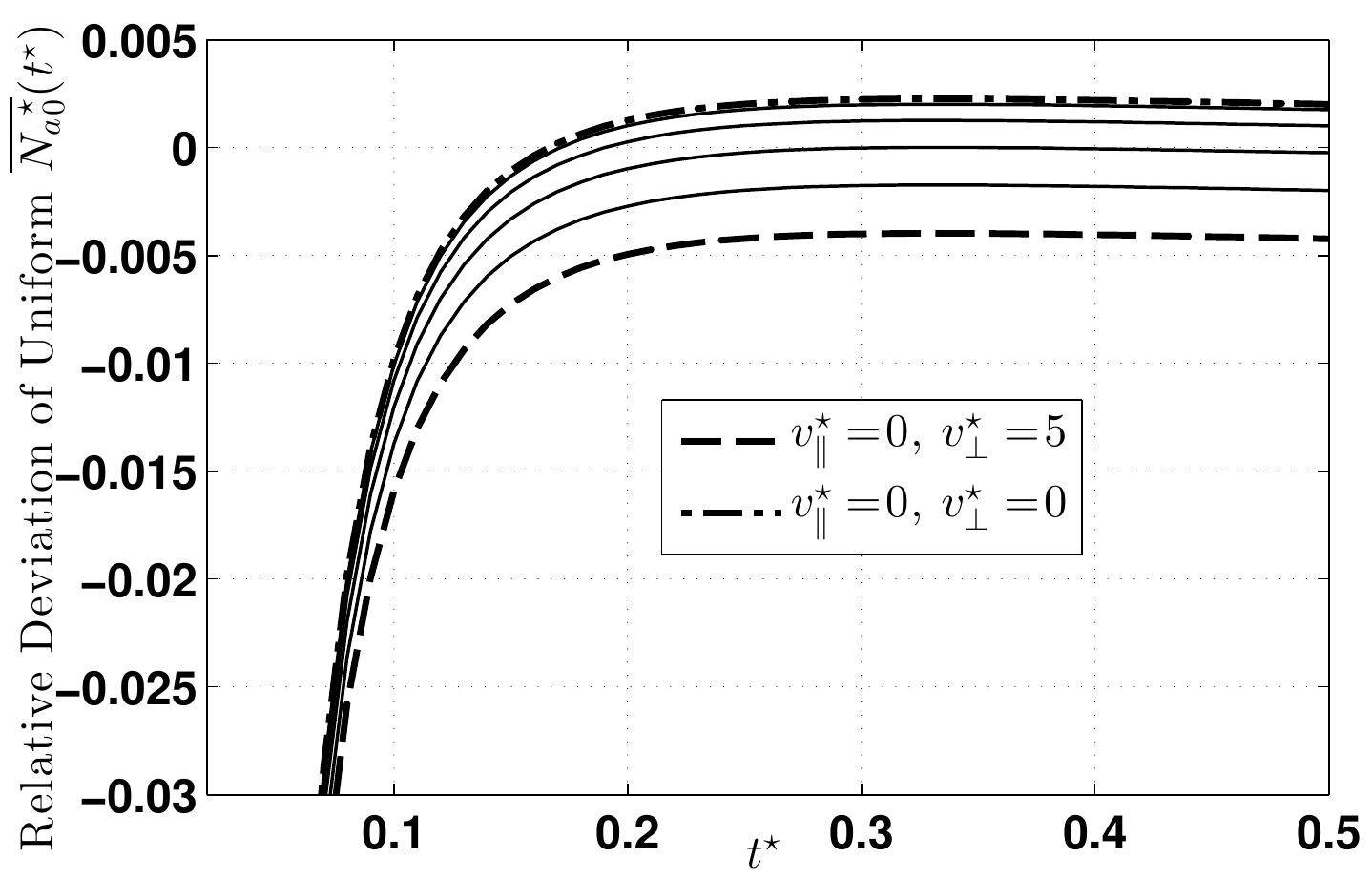}
\caption{The relative deviation in $\DMLSNxtavg{\DMLSt{}}{0}$ from the true value
(\ref{APR12_42_DMLS}) at the receiver when the uniform concentration
assumption (\ref{EQ13_07_26_DMLS}) is applied.
The flow $\Pecper$ is varied from $0$
to $5$ in increments of $1$.}
\label{uniform_vy}
\end{figure}

In Fig.~\ref{uniform_vy}, we assess the uniform concentration assumption
over time while varying $\Pecper$ from $0$ to $5$ (by symmetry,
$\DMLSNxtavg{\DMLSt{}}{0}$ due to $\Pecper < 0$ is equal to that due to
$\Pecper > 0$). Similar to $\Pecpara < 0$, which is also
a disruptive flow, a nonzero $\Pecper$ increases the
time that $\DMLSNxtavg{\DMLSt{}}{0}$
is underestimated. However, this does not significantly impact the
general accuracy of the uniform concentration assumption, as the
deviation for all flows shown is no more than $1.7\,\%$ for all
$\DMLSt{} > 0.1$.

In summary, we observe that the uniform concentration assumption
cannot be universally applied to all degrees of flow at any time
with a high degree of accuracy. The deviation of the assumption
generally increases with the magnitude of the flow, and the assumption
is least accurate immediately after the release of molecules
by the transmitter. The benefit of the assumption is analytical
tractability and simplicity in comparing the performance of detectors.
To maximize accuracy, we do not apply
the uniform concentration assumption
in the evaluation of (\ref{APR12_42_DMLS}) in the remainder of this
paper.

\subsection{Detector Performance}

Our simulations to assess detector performance as a function
of flow are executed in the particle-based stochastic framework
that we described in \cite{RefWorks:631, RefWorks:662}, and the
environment parameters are listed in Table~\ref{table_param}.
For simplicity, the $\M$ observations are
equally spaced within each bit interval.
The chosen $\Dx{\A}$ is similar to the diffusion coefficient
of many small molecules in water at room temperature
(see \cite[Ch. 5]{RefWorks:742}), and is also comparable to
that of small biomolecules in blood plasma (see \cite{RefWorks:754}).
For reference, a maximum of $3.08$ molecules is expected from a single
emission in the no-flow case $0.042\,\metre\second$
after the molecules are released. If we set $\dist = \x_0$, then
$\DMLSr{obs} = 0.1$, the dimensionless bit interval is $0.8$, and
advection $\Pec{} = 1$
translates to a steady flow of $2\, \frac{\metre\metre}{\second}$
(on the order of average capillary blood speed, from $0.1$ to
$10\, \frac{\metre\metre}{\second}$; see \cite{RefWorks:754}).
A bit interval that is (dimensionlessly) close to $1$
means that it is close to both the typical diffusion time
and advection time when $\Pec{} = 1$.

\begin{table}[!tb]
	\centering
	\caption{System parameters used for evaluating detector performance}
	{\renewcommand{\arraystretch}{1.3}
		\begin{tabular}{|l|c|c|}
		\hline
		Parameter & Symbol & Value\\ \hline
		$\#$ of molecules per emission & $\Nemit$ 	& $10^4$	\\ \hline
		Probability of binary $1$ & $\Pone$ 	& $0.5$ 	\\ \hline
		Length of transmitter sequence & $\B{}$	& $100$ bits		\\ \hline
		Bit interval time	& $\T$		& $0.2\,$ms \\ \hline
		Diffusion coefficient \cite{RefWorks:742,RefWorks:754}
		& $\Dx{\A}$ & $10^{-9}\frac{\metre^2}{\second}$ \\ \hline
		Location of transmitter & $\x_0$	& $0.5\,\mu\metre$	\\ \hline
		Radius of receiver & $\robs$	& $50\,$nm		\\ \hline
		Expected impact of noise source(s) & $\Nnoisetavg{t}$ & $1$ molecule	\\ \hline
		Simulation step size & $\Delta t$ 	& $0.5\,\mu\second$ \\ \hline
		\end{tabular}
	}
	\label{table_param}
\end{table}

In the following figures, we consider the optimal
sequence detector, the matched filter detector, and the equal weight
detector. The expected error
probabilities (evaluated using \cite[Eq. 44]{RefWorks:747})
and those found via simulation are averaged over all $100$ bits
in the transmitter sequence and averaged over $1000$ bit sequences.
The accuracy of the expected bit error probability
decreases slightly as $\M$ increases,
because the assumption that the samples are independent becomes less valid.

\begin{figure}[!tb]
\centering
\includegraphics[width=\linewidth]{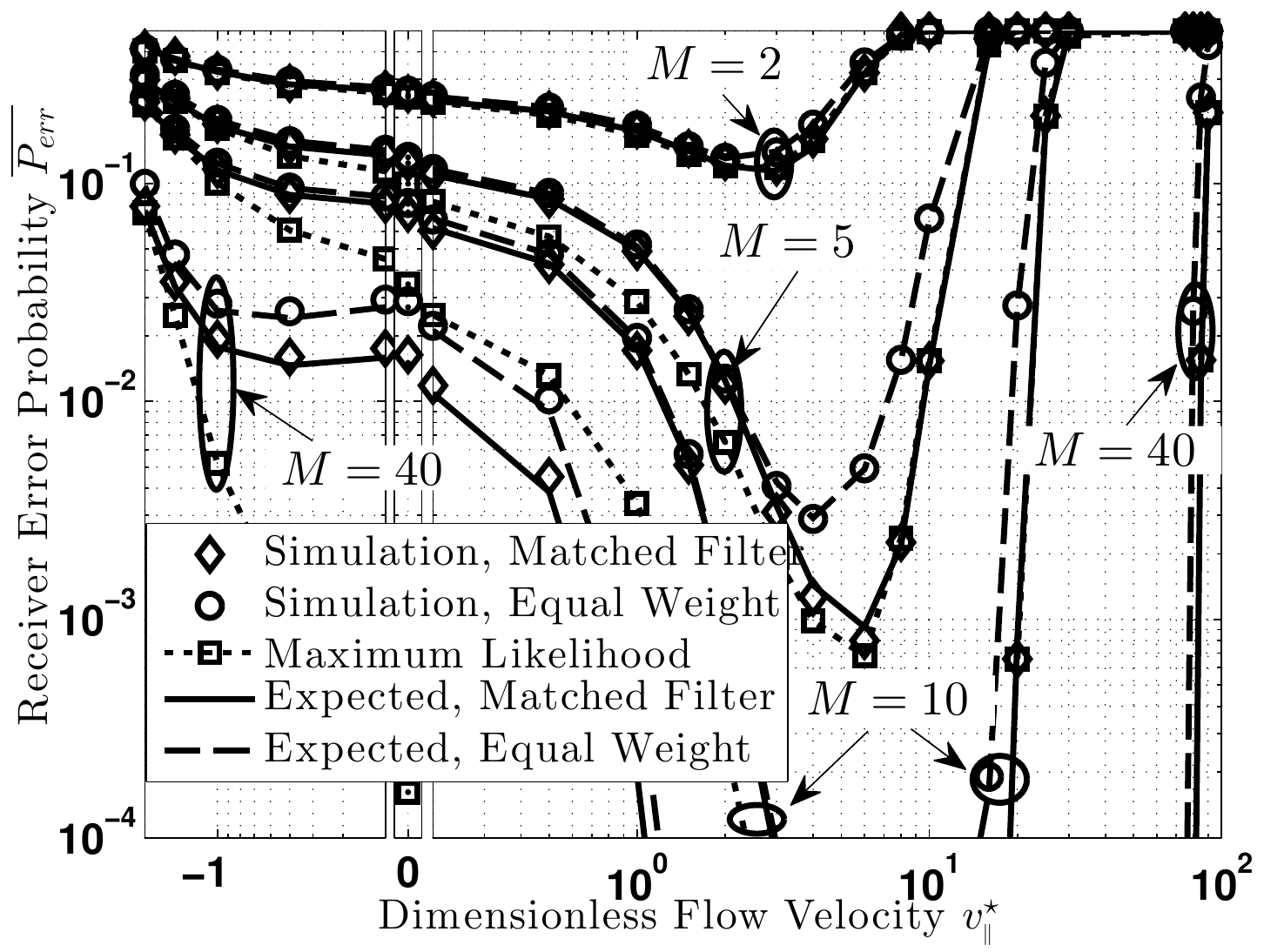}
\caption{Receiver error probability as a function of $\Pecpara$ for
$\M = \{2,5,10,40\}$ observations in each bit interval ($\Pecper = 0$).
The horizontal axis is separated into 3 regions in order to show logarithmic
scales: $-2 \le \Pecpara \le -0.2$ on a log
scale, $\Pecpara = 0$, and $0.2 \le \Pecpara \le 100$ on a log scale.}
\label{vary_vx}
\end{figure}

In Fig.~\ref{vary_vx}, we consider the impact of varying $\Pecpara$, i.e.,
the flow is either in the direction of information transmission
or directly opposite. The performance of all three detectors is quite similar
for low $\M$ and any value of $\Pecpara$, but the optimal detector
can become orders of magnitude better than the weighted sum detectors
for large $\M$, i.e., $\M > 10$.
Generally, all detectors improve
over the no-flow case when $\Pecpara > 0$. As $\Pecpara$
increases, advection dominates diffusion and the impact of \ISI~is
mitigated. However, with sufficiently high $\Pecpara > 0$,
the molecules enter and leave the receiver between consecutive
observations. As expected, communication degrades even though
the flow is actually non-disruptive.
We see this trend for $\M = 2$ when $\Pecpara> 2$,
where the advection time ($\dist/\Pecpara$, or $0.125$\;ms for $\Pecpara=
2$) becomes on the order of the time between observations ($0.1$\;ms).
This trend continues for every value of $\M$ as $\Pecpara$ increases,
but degradation would never occur as $\Pecpara \to \infty$ if the receiver
was perfectly synchronized to make an observation when the emitted molecules
pass through. We also note that, in practice, a physical receiver
would have receptors to which the emitted molecules could bind and then be
observed. However, communication under a very strong flow could still degrade if
the binding rate of the receptors was not sufficiently high.

All detectors fail when $\Pecpara$
is negative and sufficiently large,
since advection-dominant disruptive flow prevents all transmitted molecules
from reaching the receiver. However, this degradation is less severe for
small $\Pecpara$ and,
with sufficient sampling (i.e., $\M=40$), Fig.~\ref{vary_vx} shows
that both weighted sum detectors perform
\emph{better} (albeit slightly) than the no-flow case when
$-1 < \Pecpara < -0.5$.
Interestingly, the impact of this disruptive flow's removal of $\A$ molecules
in their intended bit interval is mitigated by the removal of \ISI~molecules.
Bi-directional transmission is thus possible in an environment with a steady
flow moving in a direction parallel to the line between two transceivers, as
long as advection does not dominate diffusion, and communication in each
direction can be improved for some flows over the no-flow case if weighted sum
detectors are used with a large number of samples.

\begin{figure}[!tb]
\centering
\includegraphics[width=\linewidth]{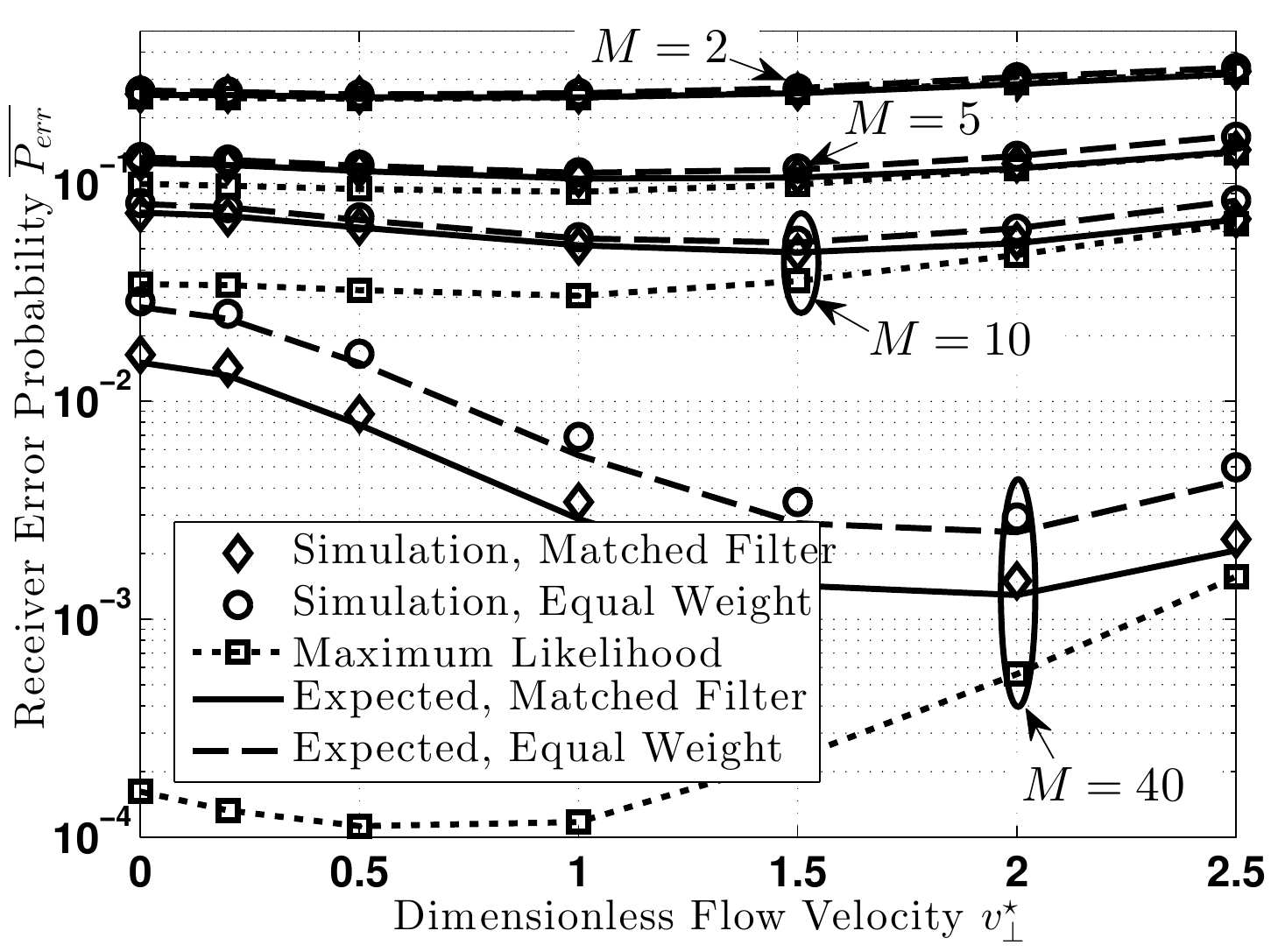}
\caption{Receiver error probability as a function of $\Pecper$ for
$\M = \{2,5,10,40\}$ observations in each bit interval ($\Pecpara = 0$).}
\label{vary_vy}
\end{figure}

In Fig.~\ref{vary_vy}, we consider the impact of varying $\Pecper$, i.e.,
the flow is perpendicular to the direction of information transmission.
As might be expected, the impact of this disruptive flow is measurably
different from $\Pecpara < 0$.
For all $\M>2$ shown, \emph{all} detectors (including the maximum likelihood
optimal sequence detector) have a range of $\Pecper$ over which they
perform better than in the no-flow case, and the potential for improvement
increases with $\M$.
As with $\Pecpara < 0$, the impact of this disruptive flow's removal of $\A$
molecules in their intended bit interval is mitigated by the removal of
\ISI~molecules, i.e., performance improves if the removal of \ISI~molecules is
proportionately greater than the degradation of the useful signal. For example,
the maximum improvement in the probability of error over the no-flow case is
about $10\,\%$ at $\Pecper=1$ when $\M = 5$, but the probability of error of
the weighted sum detectors decreases by an order of magnitude when $\M = 40$
and $1.5 \le \Pecper \le 2$. The improvement in performance of the optimal
sequence detector is small for all values of $\M$ considered, although we
expected no more than a small gain because this detector already accounts for
\ISI. As
with negative $\Pecpara$, all detectors eventually degrade as $\Pecper$
increases and advection begins to dominate diffusion.
However, the detectors do not appear to be as sensitive to $\Pecper$
as they are to the corresponding
negative values of $\Pecpara$ in Fig.~\ref{vary_vx}.
As long as a flow moving in a direction perpendicular to the line between
two transceivers does not dominate diffusion, bi-directional communication
is not only possible, but can also be improved over the no-flow case if
the detectors take enough samples.

\begin{figure}[!tb]
\centering
\includegraphics[width=\linewidth]{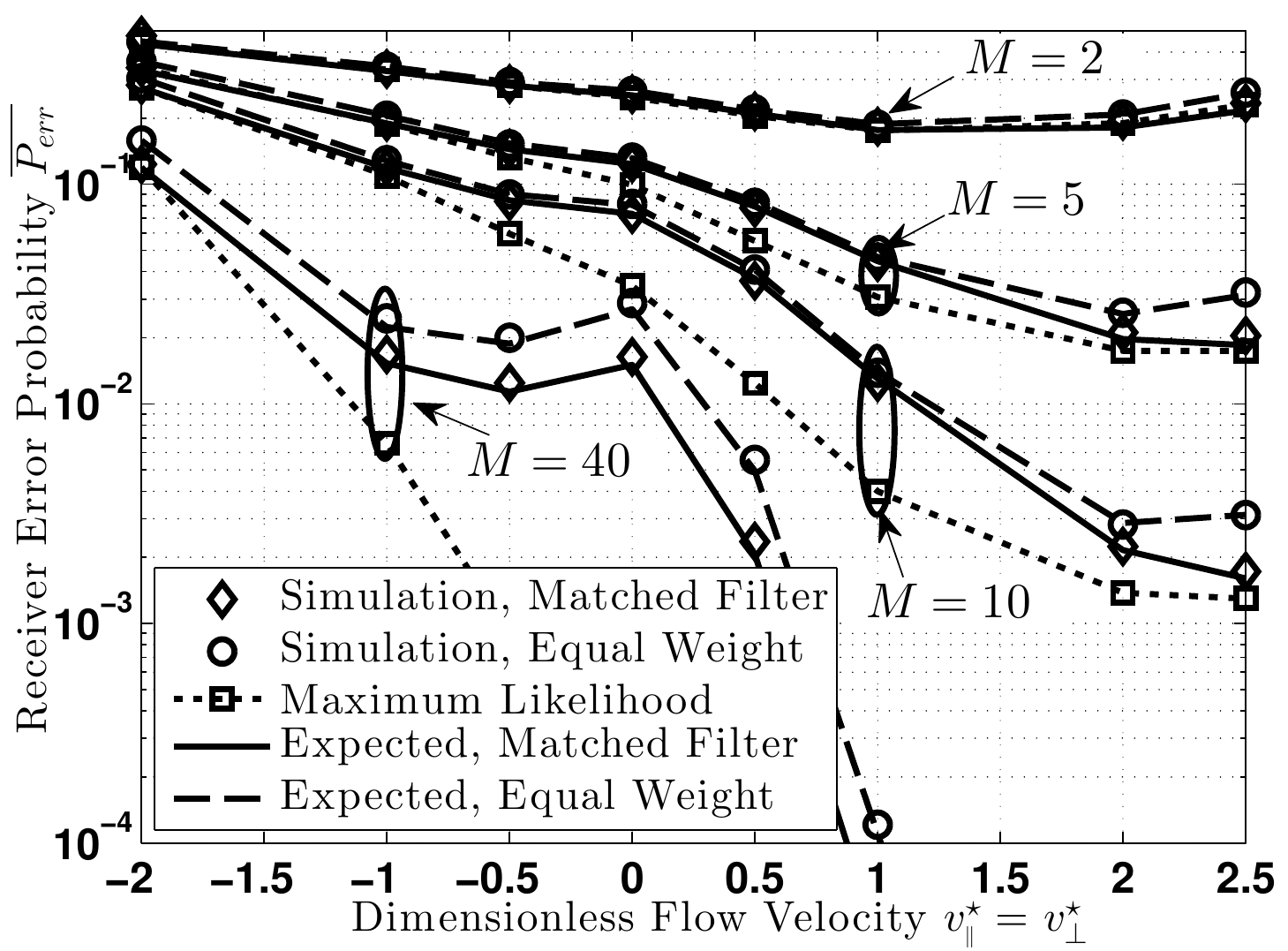}
\caption{Receiver error probability as a function of $\Pecpara = \Pecper$ for
the numbers of observations in each bit interval $\M = \{2,5,10,40\}$.}
\label{vary_vxy}
\end{figure}

Finally, in Fig.~\ref{vary_vxy} we consider the impact of varying both
$\Pecpara$ and $\Pecper$ simultaneously, such that
$\Pecpara = \Pecper$. Detector performance is most similar to that
shown in Fig.~\ref{vary_vx} because the detectors are more sensitive to
$\Pecpara$. However, there are two notable differences with
Fig.~\ref{vary_vx}. First, the improvement in the weighted sum detectors
is slightly more pronounced in Fig.~\ref{vary_vxy} when
$\M=40$ and both flows are small and negative; the expected bit error
probability of the matched filter detector is a little more than $0.01$
when $\Pecpara = \Pecper = -0.5$ but almost $0.015$ when only
$\Pecpara = -0.5$. This is an example of disruptive flows in two
dimensions contributing constructively.
Second, the degradation of all detectors occurs sooner for positive
$\Pecpara$ and $\Pecper$ than for positive $\Pecpara$ alone;
the probability of error for the equal weight detector and $\M = 5$
or $\M = 10$ begins increasing as a function of flow when
$\Pecpara = \Pecper = 2.5$, whereas it was still decreasing
when only $\Pecpara = 2.5$. This is an example of the benefits
of a non-disruptive flow component being mitigated by a disruptive
flow component.

\section{Conclusions}
\label{sec_concl}

In this paper, we studied the impact of steady uniform flow on a diffusive
molecular communication system with a passive receiver. We derived the
expected number of information molecules to be observed at the receiver
when the transmitter is any distance away. A closed-form solution is
available only when there is no flow component perpendicular to the
direction of information transmission. We showed that there are
conditions under which it is accurate to assume that the concentration of
molecules expected at the receiver is uniform, thereby simplifying
analysis.
The simulation of detector performance showed that weighted
sum detectors can perform better than in a no-flow environment when slow
disruptive flows are present, and even optimal sequence detectors can
perform slightly better than in a no-flow environment when
there is a non-dominant
disruptive flow perpendicular to the direction of information transmission.
When flows become fast enough to dominate diffusion, disruptive flows
prevent communication, whereas performance under a non-disruptive flow
is only limited by the sampling times of the detector.

\bibliography{../references/nano_ref}

\begin{thebibliography}{10}
\providecommand{\url}[1]{#1}
\csname url@samestyle\endcsname
\providecommand{\newblock}{\relax}
\providecommand{\bibinfo}[2]{#2}
\providecommand{\BIBentrySTDinterwordspacing}{\spaceskip=0pt\relax}
\providecommand{\BIBentryALTinterwordstretchfactor}{4}
\providecommand{\BIBentryALTinterwordspacing}{\spaceskip=\fontdimen2\font plus
\BIBentryALTinterwordstretchfactor\fontdimen3\font minus
  \fontdimen4\font\relax}
\providecommand{\BIBforeignlanguage}[2]{{%
\expandafter\ifx\csname l@#1\endcsname\relax
\typeout{** WARNING: IEEEtran.bst: No hyphenation pattern has been}%
\typeout{** loaded for the language `#1'. Using the pattern for}%
\typeout{** the default language instead.}%
\else
\language=\csname l@#1\endcsname
\fi
#2}}
\providecommand{\BIBdecl}{\relax}
\BIBdecl

\bibitem{RefWorks:540}
I.~F. Akyildiz, F.~Brunetti, and C.~Blazquez, ``Nanonetworks: A new
  communication paradigm,'' \emph{Computer Networks}, vol.~52, no.~12, pp.
  2260--2279, May 2008.

\bibitem{RefWorks:608}
T.~Nakano, M.~J. Moore, F.~Wei, A.~V. Vasilakos, and J.~Shuai, ``Molecular
  communication and networking: Opportunities and challenges,'' \emph{IEEE
  Trans. Nanobiosci.}, vol.~11, no.~2, pp. 135--148, Jun. 2012.

\bibitem{RefWorks:750}
G.~A. Truskey, F.~Yuan, and D.~F. Katz, \emph{Transport Phenomena in Biological
  Systems}, 2nd~ed.\hskip 1em plus 0.5em minus 0.4em\relax Pearson Prentice
  Hall, 2009.

\bibitem{RefWorks:751}
R.~B. Bird, W.~E. Stewart, and E.~N. Lightfoot, \emph{Transport Phenomena},
  2nd~ed.\hskip 1em plus 0.5em minus 0.4em\relax John Wiley \& Sons, 2002.

\bibitem{RefWorks:513}
D.~Miorandi, ``A stochastic model for molecular communications,'' \emph{Nano
  Commun. Net.}, vol.~2, no.~4, pp. 205--212, Dec. 2011.

\bibitem{RefWorks:607}
S.~Kadloor, R.~R. Adve, and A.~W. Eckford, ``Molecular communication using
  {B}rownian motion with drift,'' \emph{IEEE Trans. Nanobiosci.}, vol.~11,
  no.~2, pp. 89--99, Jun. 2012.

\bibitem{RefWorks:615}
K.~V. Srinivas, A.~W. Eckford, and R.~S. Adve, ``Molecular communication in
  fluid media: The additive inverse {G}aussian noise channel,'' \emph{IEEE
  Trans. Inf. Theory}, vol.~58, no.~7, pp. 4678--4692, Jul. 2012.

\bibitem{RefWorks:747}
\BIBentryALTinterwordspacing
A.~Noel, K.~C. Cheung, and R.~Schober, ``Optimal receiver design for diffusive
  molecular communication with flow and additive noise,'' \emph{Submitted to
  IEEE Trans. Nanobiosci.}, Jul. 2013. [Online]. Available:
  \url{arXiv:1308.0109}
\BIBentrySTDinterwordspacing

\bibitem{RefWorks:734}
H.~ShahMohammadian, G.~G. Messier, and S.~Magierowski, ``Nano-machine molecular
  communication over a moving propagation medium,'' \emph{Nano Commun. Net.},
  vol.~4, no.~3, pp. 142--153, Sep. 2013.

\bibitem{RefWorks:706}
A.~Noel, K.~C. Cheung, and R.~Schober, ``Using dimensional analysis to assess
  scalability and accuracy in molecular communication,'' in \emph{Proc. 2013
  IEEE ICC MONACOM}, Jun. 2013, pp. 818--823.

\bibitem{RefWorks:586}
J.~Crank, \emph{The Mathematics of Diffusion}, 2nd~ed.\hskip 1em plus 0.5em
  minus 0.4em\relax Oxford University Press, 1980.

\bibitem{RefWorks:630}
H.~C. Berg, \emph{Random Walks in Biology}.\hskip 1em plus 0.5em minus
  0.4em\relax Princeton University Press, 1993.

\bibitem{RefWorks:633}
T.~Szirtes, \emph{Applied Dimensional Analysis and Modeling}, 2nd~ed.\hskip 1em
  plus 0.5em minus 0.4em\relax Butterworth-Heinemann, 2007.

\bibitem{RefWorks:700}
E.~W. Ng and M.~Geller, ``A table of integrals of the error functions,''
  \emph{J. Res. Bur. Stand.}, vol. 73B, no.~1, pp. 1--20, Jan.-Mar. 1969.

\bibitem{RefWorks:773}
M.~J. Moore and T.~Nakano, ``Oscillation and synchronization of molecular
  machines by the diffusion of inhibitory molecules,'' \emph{IEEE Trans.
  Nanotechnol.}, vol.~12, no.~4, pp. 601--608, Jul. 2013.

\bibitem{RefWorks:587}
P.~Nelson, \emph{Biological Physics: Energy, Information, Life}, updated
  1st~ed.\hskip 1em plus 0.5em minus 0.4em\relax W. H. Freeman and Company,
  2008.

\bibitem{RefWorks:631}
\BIBentryALTinterwordspacing
A.~Noel, K.~C. Cheung, and R.~Schober, ``Improving diffusion-based molecular
  communication with unanchored enzymes,'' in \emph{Proc. 2012 ICST BIONETICS},
  Dec. 2012. [Online]. Available: \url{arXiv:1305.1783}
\BIBentrySTDinterwordspacing

\bibitem{RefWorks:662}
\BIBentryALTinterwordspacing
------, ``Improving receiver performance of diffusive molecular communication
  with enzymes,'' \emph{to appear in IEEE Trans. Nanobiosci.}, 2014. [Online].
  Available: \url{dx.doi.org/10.1109/TNB.2013.2295546}
\BIBentrySTDinterwordspacing

\bibitem{RefWorks:742}
E.~L. Cussler, \emph{Diffusion: Mass transfer in fluid systems}.\hskip 1em plus
  0.5em minus 0.4em\relax Cambridge University Press, 1984.

\bibitem{RefWorks:754}
A.~A. Merrikh and J.~L. Lage, ``Effect of blood flow on gas transport in a
  pulmonary capillary,'' \emph{Journal of Biomech. Eng.}, vol. 127, no.~3, pp.
  432--439, Jun. 2005.

\end{thebibliography}

\end{document}